\documentclass[12pt]{article}
\usepackage{amssymb,amsmath,cite,graphics,epsfig,subfigure}
\catcode`\@=11

\@addtoreset{equation}{section}
\def\theequation{\arabic{section}.\arabic{equation}}
     
\catcode`\@=11
\def\thesection{\arabic{section}.}

\def\appendix{\setcounter{section}{0}
        \def\thesection{Appendix.}
        \def\theequation{\Alph{section}.\arabic{equation}}}
\def\section{\@startsection{section}{1}{\z@}{3.5ex plus 1ex minus
   .2ex}{2.3ex plus .2ex}{\large\bf}}
   
\long\def\@makefntext#1{\parindent 0cm\noindent
\hbox to 1em{\hss$^{\@thefnmark}$}#1}

\newcommand{\captionfonts}{\small}
\makeatletter  
\long\def\@makecaption#1#2{%
  \vskip\abovecaptionskip
  \sbox\@tempboxa{{\captionfonts #1: #2}}%
  \ifdim \wd\@tempboxa >\hsize
    {\captionfonts #1: #2\par}
  \else
    \hbox to\hsize{\hfil\box\@tempboxa\hfil}%
  \fi
  \vskip\belowcaptionskip}
\makeatother   
\newcommand{\rvec} {\ensuremath{\vec{\mathbf r}}}
\newcommand{\rpvec}{\ensuremath{\vec{\mathbf r}\kern 1pt'}}

\begin{document}
\begin{titlepage}
\vspace{.5in}
\begin{flushright}
UCD-06-14\\
gr-qc/0606120\\
June 2006\\
\end{flushright}
\vspace{.5in}
\begin{center}
{\Large\bf
 A possible experimental test\\[.5ex]
 of quantized gravity}\\
\vspace{.4in}
{P.\ J.\ S{\sc alzman}\footnote{\it email: p@dirac.org} 
and S.~C{\sc arlip}\footnote{\it email: carlip@physics.ucdavis.edu}\\
       {\small\it Department of Physics, University of California}\\
       {\small\it Davis, CA 95616, USA}}
\end{center}

\vspace{.5in}
\begin{center}
{\large\bf Abstract}
\end{center}
\begin{center}
\begin{minipage}{4.75in}
{\small While it is widely believed that gravity should 
ultimately be treated as a quantum theory, there remains 
a possibility that general relativity should not be quantized.  
If this is the case, the coupling of classical gravity to the 
expectation value of the quantum stress-energy tensor will 
naturally lead to nonlinearities in the Schr{\"o}dinger 
equation.  By numerically investigating time evolution in 
the nonrelativistic ``Schr{\"o}dinger-Newton'' approximation, 
we show that such nonlinearities may be observable in the
next generation of molecular interferometry experiments.
}
\end{minipage}
\end{center}
\end{titlepage}
\addtocounter{footnote}{-2}

\section{Introduction}

The first attempts to quantize general relativity appeared in
the early 1930s, and in the seventy years that have followed, we
have learned a great deal about quantum field theory and gravity. 
But despite the hard work of many outstanding physicists, a 
complete, consistent theory of quantum gravity still seems distant 
\cite{Carlip}.  Given the severe difficulties, one might reasonably
ask whether the whole project could be a blind alley. Perhaps our
prejudice that everything in Nature should be quantized is simply
wrong; perhaps general relativity, a theory of spacetime, is 
fundamentally different from theories of fields within spacetime.

Somewhat surprisingly, the possibility that gravity is essentially 
classical has not yet been excluded.
The simplest model of classical general relativity coupled to 
quantum matter, sometimes called ``semiclassical gravity,'' was 
proposed forty years ago by M{\o}ller \cite{Moller} and Rosenfeld 
\cite{Rosenfeld}.  The Einstein field equations become
\begin{equation}
G_{\mu\nu} = 8\pi G \langle\psi| T_{\mu\nu}|\psi\rangle ,
\label{a1}
\end{equation}
where the operator-valued stress-energy tensor is replaced by an 
expectation value.  As a Hartree-like approximation to quantized 
gravity, such a system certainly makes sense.  But as Kibble and 
Randjbar-Daemi stressed \cite{Kibble}, viewed as a fundamental theory, 
such a model leads to nonlinearities in quantum mechanics: the 
Schr{\"o}dinger equation for the wave function $|\psi\rangle$ 
depends on the metric, which in turn depends, through (\ref{a1}), 
on $|\psi\rangle$.\footnote{Dirac was also aware of this: see 
\cite{vonB}, p.\ 1.}  While more complicated models are possible, any 
such theory must couple classical gravity to quantum sources, and it 
is hard to see how to do so without introducing similar nonlinearities.

While the literature contains a number of criticisms of semiclassical 
gravity \cite{Kibble,Duff,Eppley,Page,Unruh}, none seems decisive 
\cite{Rosenfeld,vonB,Callender,Mattingly}.  For example, one might argue 
that measurements with nonquantized gravitational waves could violate
the uncertainty principle for quantized matter \cite{Eppley}.  But
there are intrinsic limitations to the measurement of even a classical
gravitational field \cite{vonB,Smolin}; it is possible, for instance, 
that the necessary measurement would require an apparatus massive 
enough to collapse into a black hole \cite{Mattingly}.
On the experimental side, neutron interferometry \cite{COW} and 
microscopic deflection experiments \cite{Nesvizhevsky} show that 
quantum matter interacts gravitationally as expected, but these
results do not require quantization of the gravitational field itself.  
More direct experimental tests have been suggested using superpositions
in Bose-Einstein condensates \cite{Peres} or, in principle, gravitational
radiation from quantum systems \cite{Ford}, but these are not yet
practical.  The standard mechanism for the appearance of inhomogeneities
in inflationary cosmology is probably incompatible with semiclassical
gravity, requiring at least linear quantum fluctuations in the metric 
\cite{Hu}, but there are certainly other ways to produce the observed 
inhomogeneities.

Semiclassical gravity is probably excluded in a strict Everett 
interpretation of quantum mechanics \cite{Page}, but one would like
to have a result that does not depend on measurement theory or the
interpretation of quantum mechanics.  It is therefore important to 
ask whether the nonlinearities in semiclassical gravity might be 
experimentally accessible. 

The full semiclassical equations (\ref{a1}) are extremely difficult
to analyze.  But for many purposes, the nonrelativistic Newtonian 
approximation should be sufficient.  We are thus led to the
``Schr{\"o}dinger-Newton equation'' \cite{Diosi,Penrose},
\begin{equation}
i\hbar\frac{\partial\psi}{\partial t} =
-\frac{\hbar^2}{2m}{\nabla}^2\psi - m\Phi\psi ,\qquad
{\nabla}^2 \Phi = 4\pi G m|\psi|^2 ,
\label{a2}
\end{equation}
a Schr{\"o}dinger equation for matter coupled to a classical gravitational
potential that has as its source the expectation value of the mass density.
This system has been studied in the past \cite{Moroz,Bernstein,Tod,Todb},
and a fair amount is known about the lowest eigenfunctions and eigenvalues, 
but time evolution remains much more poorly understood \cite{Harrison,%
Harrisonb,Guzman}.

In the remainder of this paper, we will investigate the Schr{\"o}dinger-Newton 
evolution of an initial Gaussian wave packet.  Qualitatively, we will show 
that self-gravitation can slow or stop the spreading of the wave packet.  
We find that despite the weakness of gravity, the resulting suppression of
interference may be observable in the next generation of matter interferometry 
experiments \cite{Zeilinger}.   

\section{Setting up the problem}

Let us first consider a few analytic properties of the Schr{\"o}dinger-Newton 
equation.  Note that although the equation is nonlinear, time evolution 
preserves the norm of $\psi$, and a conserved probability current can be 
written down:
\begin{equation}
\frac{\partial}{\partial t}|\psi|^2 
  = \vec{\nabla}\cdot\left[ \frac{i\hbar}{2m}\left( 
  \psi^*\vec{\nabla}\psi - \psi\vec{\nabla}\psi^*\right)\right] .
\label{b1}
\end{equation}
Thus, as in standard quantum mechanics, we can interpret $|\psi|^2$ 
as a probability density.

We will be interested in an initial Gaussian wave function
\begin{equation}
\psi(r,0) = \left( \frac{\alpha}{\pi} \right)^{3/4} e^{-\alpha r^2 /2}
\label{b2}
\end{equation}
with width $\alpha^{-1/2}$.  In principle, we expect a two-parameter
family of solutions, labeled by $\alpha$ and $m$.  But the 
Schr{\"o}dinger-Newton equation is invariant under the rescaling
\begin{equation}
m\rightarrow \mu m ,\qquad \vec{x}\rightarrow \mu^{-3}\vec{x},\qquad 
t\rightarrow \mu^5t ,\qquad \psi\rightarrow \mu^{9/2}\psi ,
\label{b3}
\end{equation}
so if $\psi(\alpha, m; \vec{x},t)$ is a solution, so is 
$\mu^{9/2}\psi(\mu^6\alpha,\mu m;\mu^{-3}\vec{x},\mu^5t)$.  
It therefore suffices to consider a one-parameter family of solutions. 

We have investigated the dynamical Schr{\"o}dinger-Newton equation (\ref{a2})
with initial condition (\ref{b2}) numerically, using a homegrown PDE solver
specifically designed for this problem, along with associated tools to analyze
and organize the data.  Details may be found in P.\ J.\ Salzman's dissertation
\cite{diss}, and will appear in a subsequent paper.  Briefly, the
Schr\"odinger equation can be discretized by considering the action of a time
evolution operator on the known wave function at some time $t$ to yield the
unknown wave function at some later time $t'$.  For example, two common schemes
are the ``explicit method,'' which translates the wave function forward in time,
\begin{equation*}
\psi(r,\Delta t) = e^{-i\hat{H}\Delta t/\hbar} \psi(r,0),
\end{equation*}
and the ``implicit method,'' which translates the wave function backward in
time,
\begin{equation*}
e^{+i\hat{H}\Delta t/\hbar} \psi(r,\Delta t) = \psi(r,0).
\end{equation*}
Here $\hat{H}$ is the Schr\"odinger-Newton Hamiltonian, which involves
both derivatives and the integral
\begin{equation}
I_{SN} = \iiint\limits_{\hbox{\scriptsize all space}}
  \frac{|\psi(\rpvec,t)|^2}{|\rvec-\rpvec|} d^3r',
\label{p1}
\end{equation}
the solution to the Poisson equation (\ref{a2}).  With either scheme, one 
can perform a Taylor expansion of the energy operators to any desired order,  
requiring successively higher order derivatives that can be approximated 
and converted into a numerical algorithm.

Each of these schemes, however, has properties which make it undesirable to 
use for a numerical PDE solver.  The explicit scheme, while simple and 
calculationally inexpensive, can be shown to be numerically unstable.  The 
implicit scheme, while numerically stable, requires that we find an inverse
operator, which is complicated and calculationally expensive.  Additionally, 
neither scheme is unitary.

We chose to use Cayley's form \cite{Cayley}, which is an average of the 
explicit and implicit methods:
\begin{equation}
e^{i\hat{H}\Delta t/2\hbar} \psi(r,\Delta t/2)  
  =  e^{-i\hat{H}\Delta t/2\hbar}\psi(r,-\Delta t/2) .
\label{p2}
\end{equation}
Cayley's form is unitary and is of higher order than either the implicit or
explicit method.  The evolution operator on either side of Cayley's form was
Taylor expanded to first order, giving a second order accurate algorithm.
Derivative terms were numerically approximated, and yielded a tridiagonal
system of linear equations from which the wavefunction at the next timestep
could be computed.  The integral $I_{SN}$ technically involves the wave
function at the ``current timestep,'' at which it is unknown; we linearized 
the integral by using the known wave function at the previous timestep.  
Accuracy was ensured by successively decreasing the timestep $\Delta t$ 
and looking at the limiting behavior as $\Delta t$ approached machine 
precision.

A wide assortment of techniques was employed to further test the accuracy
of our method.  As one important benchmark, for each set of input parameters 
the program was rerun with the potential turned off, and the results were 
compared with the exact analytic solution for the free particle; the 
differences were found to be negligible.

\section{Numerical Results}

The Schr{\"o}dinger-Newton equation was repeatedly solved using a 
fixed wave packet width $\alpha = 5\times10^{16}\,\mathrm{m}^{-2}$ 
while varying the mass $m$.  The results may be summarized as follows
(with masses in unified atomic mass units):
\begin{enumerate}
\item For small masses ($<4.5\,\mathrm{u}$), the behavior is essentially 
  indistinguishable from that of a free particle.  As $m$ increases, the 
  wave packet spreads more slows, as one might expect from ``self-gravitation.''
\item For masses between $1.9\times10^3\,\mathrm{u}$ and $7.2\times10^3\,
  \mathrm{u}$, the behavior is complex; the wave packet typically fluctuates
  rapidly and develops growing oscillations (figure 1a).  This seems to be
  related to the behavior found in \cite{Harrison,Harrisonb,Guzman}.
\item For masses between $7.8\times10^3\,\mathrm{u}$ and $2.9\times10^{14}\,
  \mathrm{u}$, the wave packet ``collapses,'' shrinking in width (figure 1b).
\item For larger masses, the wave packet appears stationary; we have not 
  been able to run the program long enough to determine the behavior.
\end{enumerate}
\begin{figure}[t]
\centering\mbox{\subfigure[complex]{\epsfig{figure=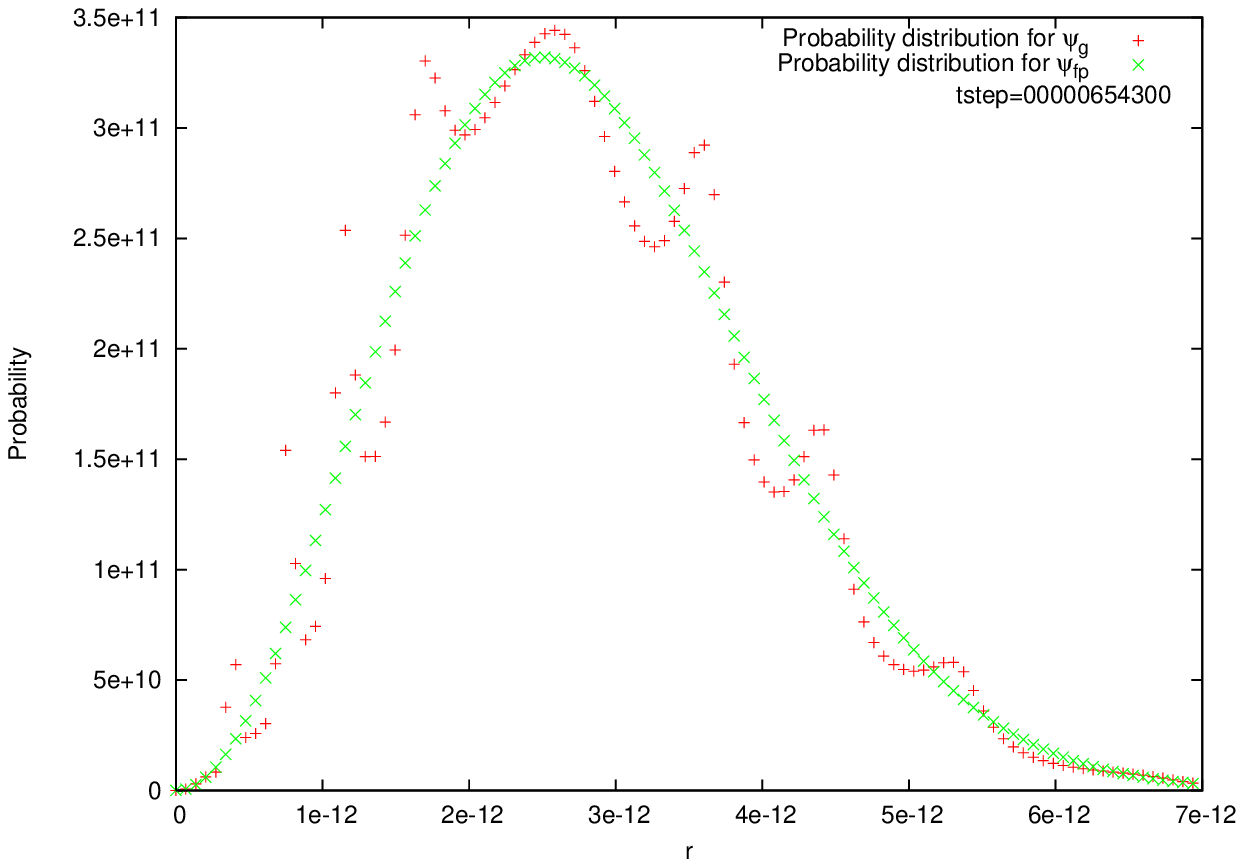,width=.4\textwidth}}%
\quad\subfigure[collapse]{\epsfig{figure=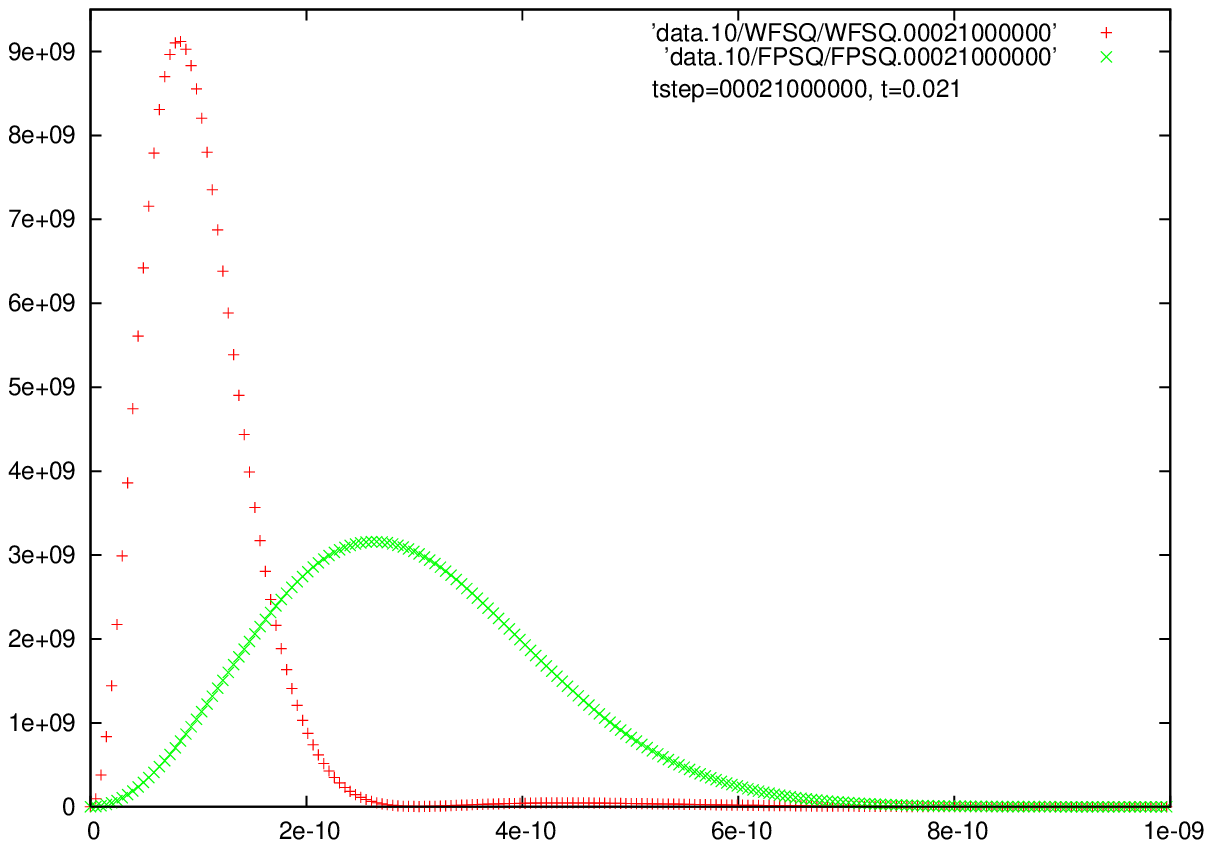,width=.4\textwidth}}}
\caption{Comparison of Schr{\"o}dinger-Newton and free particle wave functions.
For intermediate masses (a), the evolution becomes complex; for larger values 
(b), the wave function ``collapses.''}
\label{fig1}
\end{figure}

One can obtain a crude estimate of the ``collapse'' mass by noting
that for a free particle, the peak probability density of a Gaussian 
wave packet occurs at 
$r_p \sim \alpha^{-1/2}\left(1+\frac{\alpha^2\hbar^2}{m^2}t^2\right)^{1/2}$,
``accelerating'' at a speed ${\Ddot r}_p \sim \hbar^2/m^2r_p{}^3$.  Equating
this with the acceleration due to gravity at $t=0$ yields a mass
\begin{equation}
m \sim \left(\frac{\hbar^2\sqrt{\alpha}}{G}\right)^{1/3} 
 \sim 10^{10}\,\mathrm{u},
\label{d0}
\end{equation}
lying roughly at the middle of the mass range for which we see ``collapse.''
This expression behaves properly under the scaling (\ref{b3}), and could be
guessed by dimensional analysis; a key result of our numerical simulations 
is that onset of ``collapse'' occurs at significantly smaller masses, 
presumably reflecting the nonlinearity of the evolution.

Using the scaling (\ref{b3}), we can summarize our results by the ``phase
diagram'' shown in figure 4.  Of particular interest for experiment is the 
``collapsing'' phase (C).  As the figure illustrates, the wave packet of fixed  
width $w=\alpha^{-1/2}$ will shrink in width if its initial mass lies 
within a range $m_-(w)<m<m_+(w)$.  Numerically, with $w$ in microns and $m$ 
in unified atomic mass units, we find
\begin{equation}
m_-/1\,\mathrm{u} = 1300({w/1\,\mathrm{\mu m}})^{-1/3}, 
  \qquad
m_+/1\,\mathrm{u} = 4.8\times10^{13}({w/1\,\mathrm{\mu m}})^{-1/3}. 
\label{d1}
\end{equation}
The characteristic collapse times in nanoseconds, obtained numerically and 
scaled according to (\ref{b3}), are
\begin{equation}
T_-/1\,\mathrm{ns} = 1.2\times10^{-4}({w/1\,\mathrm{\mu m}})^{-5/3}
  \qquad
T_+/1\,\mathrm{ns} = 1.2\times10^{-2}({w/1\,\mathrm{\mu m}})^{-5/3}
\label{d2}
\end{equation}
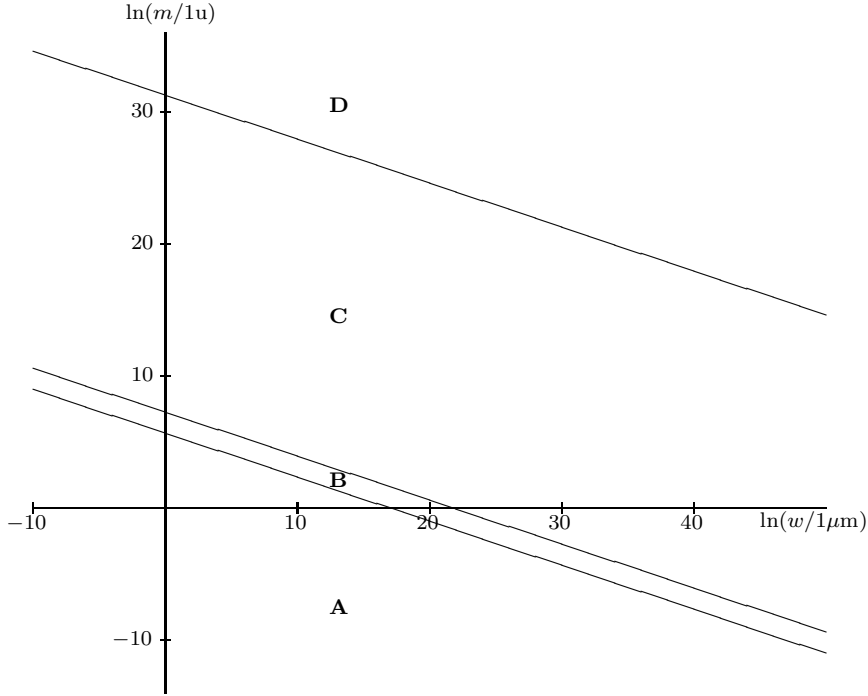
\begin{figure}
\centering
\begin{picture}(300,200)(0,40)
\put(0,100){\line(1,0){300}}
\put(50,30){\line(0,1){250}}
\multiput(0,98)(50,0){6}{\line(0,1){4}}
\multiput(48,50)(0,50){5}{\line(1,0){4}}
\put(275,93){$\scriptstyle \ln(w/1\mathrm{\mu m})$}
\put(35,285){$\scriptstyle \ln(m/1\mathrm{u})$}
\put(0,145){\line(3,-1){300}}
\put(0,153){\line(3,-1){300}}
\put(0,273){\line(3,-1){300}}
\put(-10,92){$\scriptstyle -10$}
\put(95,92){$\scriptstyle 10$}
\put(145,92){$\scriptstyle 20$}
\put(195,92){$\scriptstyle 30$}
\put(245,92){$\scriptstyle 40$}
\put(30,48){$\scriptstyle -10$}
\put(37,148){$\scriptstyle 10$}
\put(37,198){$\scriptstyle 20$}
\put(37,248){$\scriptstyle 30$}
\put(112,60){$\scriptstyle\mathbf{A}$}
\put(112,108){$\scriptstyle\mathbf{B}$}
\put(112,170){$\scriptstyle\mathbf{C}$}
\put(112,250){$\scriptstyle\mathbf{D}$}
\end{picture}
\caption{``Phase diagram'' for Schr{\"o}dinger-Newton solutions.  Region A:
wave packets spread; region B: complex behavior; region C: collapse; region D:
undetermined by our simulation.}
\label{fig2}
\end{figure}
\vspace*{-4ex}
 
\section{Experimental tests}

A ``collapsing'' wave packet will lead to a suppression of interference,
which should in principle be observable.  Most recent experiments in 
matter-wave diffraction have used Talbot-Lau interferometry \cite{Talbot}, 
in which an image of a diffraction grating with slit spacing $d$ appears at 
a distance $L_T = d^2/\lambda$ from the grating.   The heaviest molecule
for which interference has been observed to date is fluorofullerene,
C${}_{60}$F${}_{48}$, with a mass of 1632 u \cite{Hackermuller}.  The 
grating slits in this experiment have a width $w\sim.5\,\mathrm{\mu m}$.  
From (\ref{d1}), semiclassical gravity would predict a loss of interference 
for a wave packet of this width for masses greater than $m\sim 1600\,\mathrm{u}$; 
C${}_{60}$F${}_{48}$ lies just at the edge of this range.  

This is, of course, an oversimplification: the molecular wave packets in 
\cite{Hackermuller} are not spherically symmetric Gaussians.  The grating 
slits limit the width in one direction, say $y$, while the widths in the 
$x$ and $z$ directions are determined by other factors.  Repeating the
argument that led to (\ref{d0}) for a cylinder and a slab, we might expect 
a new limiting mass on the order of
\begin{equation}
m_-' = m_- (w_x/w)^{1/3}(w_z/w)^{1/3}
\label{e1}
\end{equation}
In the fluorofullerene experiment, $w_z$ was controlled by a height limiter
with a width of $100\,\mathrm{\mu m}$, giving a factor of about $6$ in (\ref{e1}).
The appropriate value for the width in the direction of the beam is less 
clear.  As a pessimistic estimate, we note that the distance from the first 
grating, which is responsible for the transverse coherence of the beam, and 
the second grating, responsible for the interference, was $.38\,\mathrm{m}$; 
using this value for $w_x$ in (\ref{e1}) would give a factor of about $90$.
Thus while the results for fluorofullerene interference probably do not
directly probe semiclassical gravity, they appear to come within two to
three orders of magnitude of a real test.

To proceed further, we need both experimental and theoretical work.  On the
theory side, simulations should be done with more realistic wave packet
profiles.  It is also important to see how sensitive our results are to
the exact (Gaussian) form of the initial wave function.  In particular,
since a stable spherically symmetric ground state exists \cite{Harrison},
\emph{some} initial profiles must have different behavior than what we 
have seen. 
 
On the experimental 
side, some progress can come from using narrower slits and from controlling 
the longitudinal width $w_x$, perhaps with a shutter to restrict the time 
each molecule enters the apparatus.  The most useful gain, though, would 
probably come from measurements of higher mass molecules.  This is not 
an unreasonable goal: experimentalists have argued that by using optical 
gratings, it may be possible to exhibit interference for molecules with masses 
of up to $10^6\,\mathrm{u}$ \cite{Zeilinger,Talbot,Hornberger,Zeilingerb}.  
If the next generation of matter wave interferometers can come even close 
to this limit, an unambiguous test of semiclassical gravity should be 
possible.

\vspace{1.5ex}
\begin{flushleft}
\large\bf Acknowledgments
\end{flushleft}

P.\ J.\ S.\ would like to thank John Robert Burke for valuable discussions.
S.\ C.\ would like to thank many colleagues at Peyresq 11, including Brandon
Carter, Larry Ford, Bei Lok Hu, Seif Randjbar-Daemi, and Albert Roura, 
for a number of useful comments and suggestions.  This work was supported 
in part by U.S.\ Department of Energy grant DE-FG02-91ER40674.

\end{document}